# Extraction of Freshwater and Energy from Atmosphere

## Alexander Bolonkin
C&R, 1310 Avenue R, #F-6, Brooklyn, NY 11229, USA
T/F 718-339-4563, aBolonkin@juno.com, http://Bolonkin.narod.ru

## Abstract

Author offers and researches a new, cheap method for the extraction of freshwater from the Earth's atmosphere. The suggested method is fundamentally dictinct from all existing methods that extract freshwater from air. All other industrial methods extract water from a saline water source (in most cases from seawater). This new method may be used at any point in the Earth except Polar Zones. It does not require long-distance freshwater transportation. If seawater is not utilized for increasing its productivity, this inexpensive new method is very environment-friendly. The author's method has two working versions: (1) the first variant the warm (hot) atmospheric air is lifted by the inflatable tube in a high altitude and atmospheric steam is condenced into freswater: (2) in the second version, the warm air is pumped 20-30 meters under the sea-surface. In the first version, wind and solar heating of air are used for causing air flow. In version (2) wind and propeller are used for causing air movment.

   The first method does not need energy, the second needs a small amount. Moreover, in variant (1) the freshwater has a high pressure (>30 or more atm.) and can be used for production of energy such as electricity and in that way the freshwater cost is lower. For increasing the productivity the seawater is injected into air and solar air heater may be used. The solar air heater produces a huge amount of electricity  as a very powerful electricity generation plant. The offered electricity installation in 100  -  200 times cheaper than any common electric plant of equivalent output.

## Introduction

   A **Freshwater** body contains low concentrations of dissolved salts and other total dissolved solids. It is an important renewable resource, necessary for the survival of most terrestrial organisms, and required by humans for drinking and agriculture, among many other uses.

Freshwater can be defined as water with less than 0.5 parts per thousand dissolved salts. Freshwater bodies include lakes, rivers, and some bodies of underground water. The ultimate source of fresh water is the precipitation of atmosphere in the form of rain and snow.

Access to unpolluted fresh water is a critical issue for the survival of many species, including humans, who must drink fresh water in order to survive. Only three percent of the water on Earth is freshwater in nature, and about two-thirds of this is frozen in glaciers and polar ice caps. Most of the rest is underground and only 0.3 percent is surface water. Freshwater lakes contain seven-eighths of this fresh surface water. Swamps have most of the balance with only a small amount in rivers.

It is estimated that 15% of world-wide water use is for household purposes. These include drinking water, bathing, cooking, sanitation, and gardening. Basic household water requirements at around 50 liters per person per day, excluding water for gardens.

Many contries and regions do not have enough freshwater.
**Desalination** refers to any of several processes that remove the excess salt and other minerals from water in order to obtain fresh water suitable for animal consumption or irrigation, and if almost all of the salt is removed for human consumption, sometimes the process produces table salt as a by-product. Desalination of ocean water is common in the Middle East (because of water scarcity) and the Caribbean, and is growing fast in the USA, North Africa, Singapore, Spain, Australia and China.

Desalination of brackish water is done in the United States in order to meet treaty obligations for river water entering Mexico. Several Middle Eastern countries have energy reserves so great that they use desalinated water for agriculture. Saudi Arabia's desalination plants account for about 24% of total world capacity.

There are a lot of methods for desacilation: Distillation, Evaporation/condensation, Multiple-effect. Membrane processes, Electrodialysis reversal, Nanofiltration, Freezing, Solar humidification, Methane hydrate crystallisation, vacuum distillation, and so on. All request a lot of energy and produce high cost freshwater.

As of July 2004, the two leading methods were Reverse Osmosis (47.2% of installed capacity world-wide) and Multi Stage Flash (36.5%).

**Reverse Osmosis.** In the last decade, membrane processes have grown very fast, and Reverse Osmosis (R.O.) has taken nearly half the world's installed capacity. Membrane processes use semi-permeable membranes to filter out dissolved material or fine solids. The systems are usually driven by high-pressure pumps, but the growth of more efficient energy-recovery devices has reduced the power consumption of these plants and made them much more viable; however, they remain energy intensive and, as energy costs rise, so will the cost of R.O. water.

The membranes used for reverse osmosis have a dense barrier layer in the polymer matrix where most separation occurs. In most cases the membrane is designed to allow only water to pass through this dense layer while preventing the passage of solutes (such as salt ions). This process requires that a high pressure be exerted on the high concentration side of the membrane, usually 2–17 bar (30–250 psi) for fresh and brackish water, and 40–70 bar (600–1000 psi) for seawater, which has around 24 bar (350 psi) natural osmotic pressure which must be overcome.

This process is best known for its use in desalination (removing the salt from sea water to get fresh water), but has also purified naturally occurring water for medical, industrial process and rinsing applicaions since the early 1970s.

**Multi-stage flash distillation** is a desalination process that distills sea water by flashing a portion of the water into steam in multiple stages. First, the seawater is heated in a container known as a brine heater. This is usually achieved by condensing steam on a bank of tubes carrying sea water through the brine heater. Thus heated, the water is passed to another container known as a "stage", where the surrounding pressure is lower than that in the brine heater. It is the sudden introduction of this water into a lower pressure "stage" that causes it to boil so rapidly as to flash into steam. As a rule, only a small percentage of this water is converted into steam. Consequently, it is normally the case that the remaining water will be sent through a series of additional stages, each possessing a lower ambient pressure than the previous "stage". As steam is generated, it is condensed on tubes of heat exchangers that run through each stage.

**Cogeneration.** There are circumstances in which it may be possible to use the same energy more than once. With cogeneration this occurs as energy drops from a high level of activity to an ambient level. Distillation processes, in particular, can be designed to take advantage of co-generation. In the Middle East and North Africa, it has become fairly common for dual-purpose facilities to produce both electricity and water. The main advantage being that a combined facility can consume less fuel than would be needed by two separate facilities.

**Economics.** A number of factors determine the capital and operating costs for desalination: capacity and type of facility, location, feed water, labor, energy, financing and concentrate disposal.

Desalination stills now control pressure, temperature and brine concentrations to optimize the water extraction efficiency. Nuclear-powered desalination might be economical on a large scale, and there is a pilot plant in the former USSR.

Critics point to the high costs of desalination technologies, especially for poor third world countries, the impracticability and cost of transporting or piping massive amounts of desalinated seawater throughout the interiors of large countries, and the "lethal byproduct of saline brine that is a major cause of marine pollution when dumped back into the oceans at high temperatures". While noting that costs are falling, and generally positive about the technology for affluent areas that are proximate to oceans, one study argues that "Desalinated water may be a solution for some water-stress regions, but not for places that are poor, deep in the interior of a continent, or at high elevation. Unfortunately, that includes some of the places with biggest water problems. Indeed, one needs to lift the water by 2000 m, or transport it over more than 1600 km to get transport costs equal to the desalination costs. Thus, desalinated water is only really expensive in places far from the sea, like New Delhi, or in high places, like Mexico City. Desalinated water is also expensive in places that are both somewhat far from the sea and somewhat high, such as Riyadh and Harare. In other places, the dominant cost is desalination, not transport. This leads to relatively low costs in places like Beijing, Bangkok, Zaragoza, Phoenix, and, of course, coastal cities like Tripoli.

**Environmental.** Regardless of the method used, there is always a highly concentrated waste product consisting of everything that was removed from the created "fresh water". These concentrates are classified by the U.S. Environmental Protection Agency as industrial wastes. With coastal facilities, it may be possible to return it to the sea without harm if this concentrate does not exceed the normal ocean salinity gradients to which osmoregulators are accustomed. Reverse osmosis, for instance, may remove 50% or more of the water, doubling the salinity of ocean waste.

The hypersaline brine has the potential to harm ecosystems, especially marine environments in regions with low turbidity and high evaporation that already have elevated salinity. Examples of such locations are the Persian Gulf, the Red Sea and, in particular, coral lagoons of atolls and other tropical islands around the world. Because the brine is more dense than the surrounding sea water due to the higher solute concentration, discharge into water bodies means that the ecosystems on the bed of the water body are most at risk because the brine sinks and remains there long enough to damage the ecosystems. Careful re-introduction attempts to minimize this problem.

The benthic community cannot accommodate such an extreme change and many filter-feeding animals are destroyed when the water is returned to the ocean. This presents an increasing problem further inland, where one needs to avoid ruining existing fresh water supplies such as ponds, rivers and aquifers. As such, proper disposal of "concentrate" needs to be investigated during the design phase.

**Experimental techniques and other developments.** In the past many novel desalination techniques have been researched with varying degrees of success. Some are still on the drawing board now while others have attracted research funding. For example, to offset the energetic requirements of desalination, the U.S. Government is working to develop practical solar desalination.

Other approaches involve the use of geothermal energy. An example would be the work being done by SDSU CITI International Consortium for Advanced Technologies and Security. From an environmental and economic point of view, in most locations geothermal desalination can be preferable to using fossil groundwater or surface water for human needs, as in many regions the available surface and groundwater resources already have long been under severe stress.

About 577,000 $km^3$ water vaporazes from Earth's surface in one years (505,000 $km^3$ of them from oceans). The one man spent from 10 - 50 liters per day (last number for industrial counries and includes the watering of house garden). The huge amount of water request the plants. For example, the one hectare of wheat requests 2000 kL (kL is kiloliter = 1 ton), the cabbage - 8000 kL, the forest 12,000 - 15,000 kL per summer.

Author offers the new cheap method for extraction freshwater from atmosphere and incidentally the extraction of energy. This method may be used in any point of Earth except Polar regions. It does not request a long distance transportation. If you not use a seawater for increasing its productivity, this method very frendly to environment.

The closed works and information about this topic the reader find in [1]-[21].

## Description of Innovations

**1. High height tube extractor and electric plant.** The offered extractor is shown in Fig.1. Main part is cheap inflatable high altitude tube 1 (up 3 - 5 km) supported by bracing wires 14. Tube is designed from inflatable toroids 13. They keep the tube form. Tube has freshwater collector 2 and freshwater pipe line 3 inside. The tube can has a film solar heater 8 (optional) located on ground. That is transparent film over ground (black surface). The entrance air flows between film and ground and is heated by solar radiation. That strong increase the speed of air flow (see computation). If we additional inject the saline water 7 into air flow, we additional increase the water productivity (see computation section). The installation can have a propeller 15 for increases air flow in windless weather. The top end of tube has air turbine and electric generator 19. The top end can also has the wind turbine 20 (optinal). The top end has also the observation desk 16 and elevator 18. The tube enter and exit 10 (fig.1) have wind leafs (2, 4 fig.2). The other parts are noted under figs. 1-2.

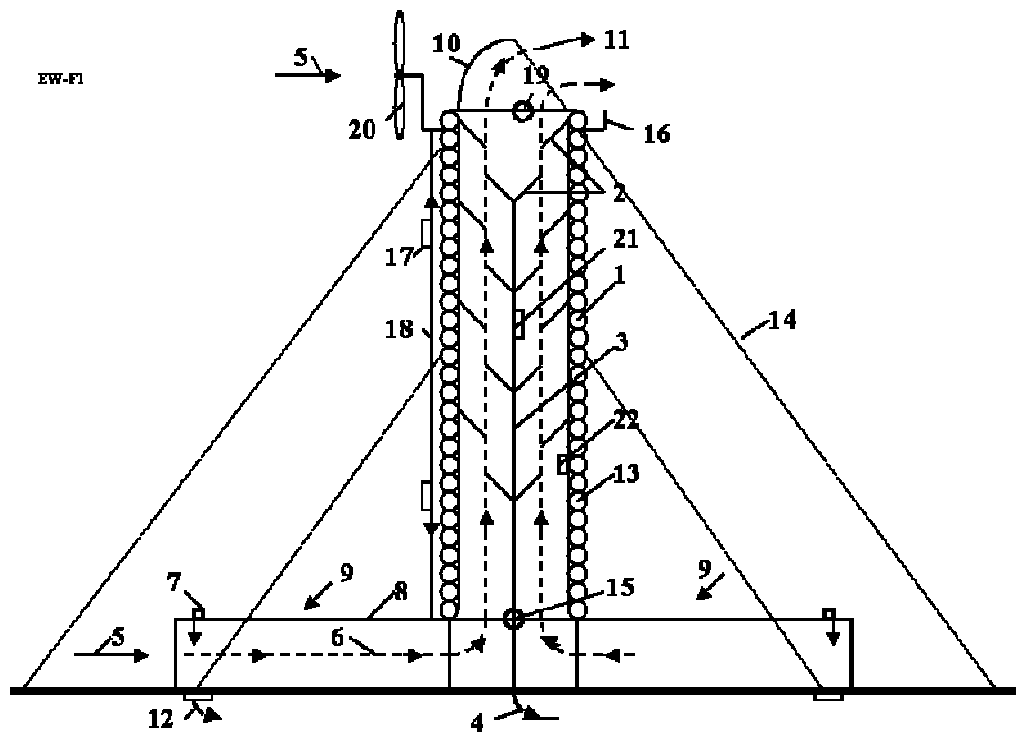

**Fig.1**. Inflatable extractor freshwater from atmosphere and electric plant (side view). Notation: 1 - vertical tube, 2 - freshwater collector, 3 - freshwater pipe line, 4 - exit of freshwater, 5 - wind, 6 - air flow, 7 - injector of sea (saline) water (optional), 8 - transparence film and solar heater of air (optional), 9 - solar radiation (optional), 10 - air exit. 11 - air flow, 12 - collector of seawater (optional), 13 - inflatable toroid, 14 - support cable (bracing wire), 15 - ventilator (propeller) (optional), 16 - observation desk for tourists and communication installation (optional), 17 - passenger cabin, 18 - elevator, 19 - top propeller-electric generator, 20 - wind electric generator, 21, 22 - mobile cabins.

The installation works the following way. The wind leafs (3, 4 fig.2) automatically are opened so to use the wind dynamic pressure (and to draw off for top end of tube). The air entrances in the solar heater 8 (or in entrance of tube when the solar heater is absent), is warmed and go to vertical tube. At high altitude the air expands, cools and air steam condenses in water collector. The pipe line delivery the freshwater under high pressure (because the altitude is big) to need region. The high pressure of water may be used for production of energy (electricity). Moreover, we can install an air turbine and electric generator at top of tube and get electricity when we do not need a big amount of freshwater. In this case the solar energy of the solar heater is transferred in electric energy (see computation).

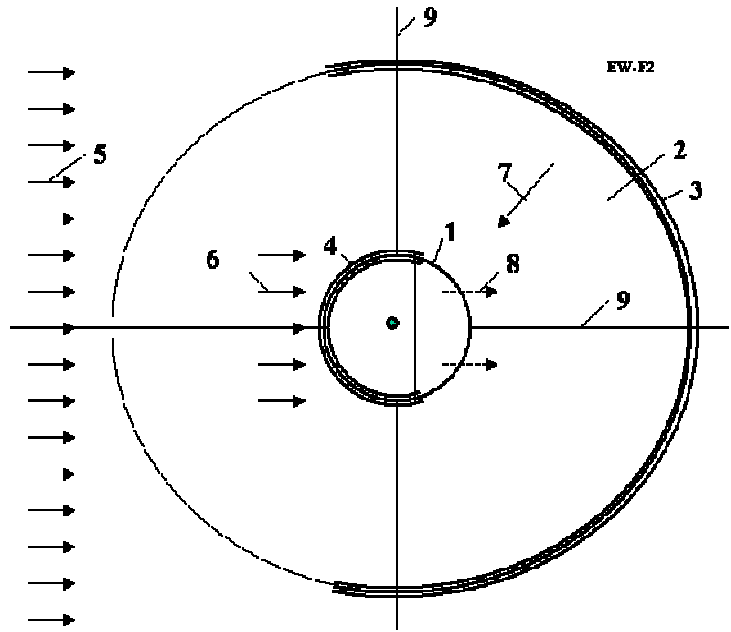

**Fig.2.** Inflatable extractor freshwater from atmosphere (top view). Notation: 1 - vertical inflatable tube, 2 - solar heater (optional), 3 - wind leafs of solar heat, 4 - wind leafs of air exit, 5 - wind at ground, 6 - wind at altitude, 7 - solar radiation, 8 - exit tube air flow, 9 - tube support cable.

**2. Sea air-water extractor.** The sea air-water extractor is shown in fig.3. That has the air wind dynamic entrance and exit which are same with previous version. New is a sea heat exchanger. That locates at depth 20-30 meters where temperature is 5 - 10 °C. It is made from steel tubes which can keep 2-3 atm outer pressure. For increasing productivity the installation has propeller (pump) which moves the air through tube and seawater injector. That can also have the solar air heater same the previous variant.

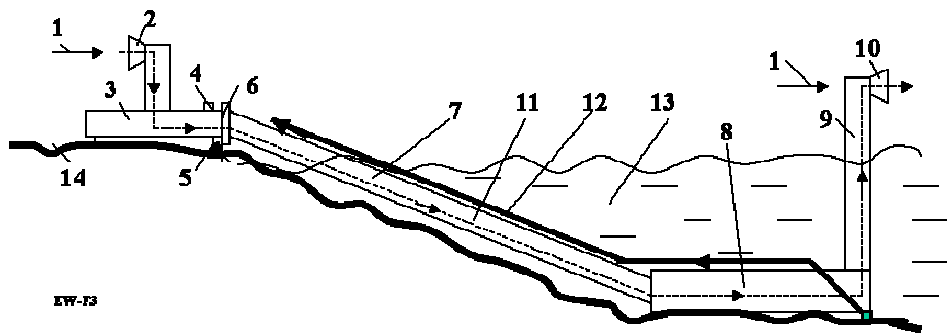

**Fig.3.** Sea extractor freshwater from atmosphere. Notation: 1 - wind, 2 - air entrance, 3 - solar air heater, 4 - injector of sea water, 5 - collector of superfluous sea water, 6 - air ventilator, 7 - air tube, 8 - radiator (heat exchanger), 9 - exit air tube, 10 - exit air flow, 11 - air flow, 12 - freshwater line, 13 - sea, 14 - ground.

# Computations and Estimations

A reader can get the equations in below from well-known physical laws. That way the author does not give the detail explanations.

1. **Amount of water in atmosphere**. Amount of water in atmosphere depends from temperature and humidity. For relative humidity 100% the maximum partial pressure of water vapor is shown in Table 1.

   **Table 1**. Maximum partial pressure of water vapor in atmosphere via air temperature

   | t, C   | -10   | 0     | 10   | 20   | 30   | 40   | 50   | 60   | 70   | 80   | 90   | 100 |
   |--------|-------|-------|------|------|------|------|------|------|------|------|------|-----|
   | p,kPa  | 0.287 | 0.611 | 1.22 | 2.33 | 4.27 | 7.33 | 12.3 | 19.9 | 30.9 | 49.7 | 70.1 | 101 |

   The amount of water in 1 m3 of air may be computed by equation
   $$m_W = 0.00625\ [p(t_2)h - p(t_1)], \qquad (1)$$
   where $m_W$ is mass of water, kg in 1 m$^3$ of air; $p(t)$ is vapor (steam) pressure from Table 1, relative $h = 0 \div 1$ is relative humidity. The computation of equation (1) is presented in fig.4. Typical relative humidity of atmosphere air is 0.5 - 1.

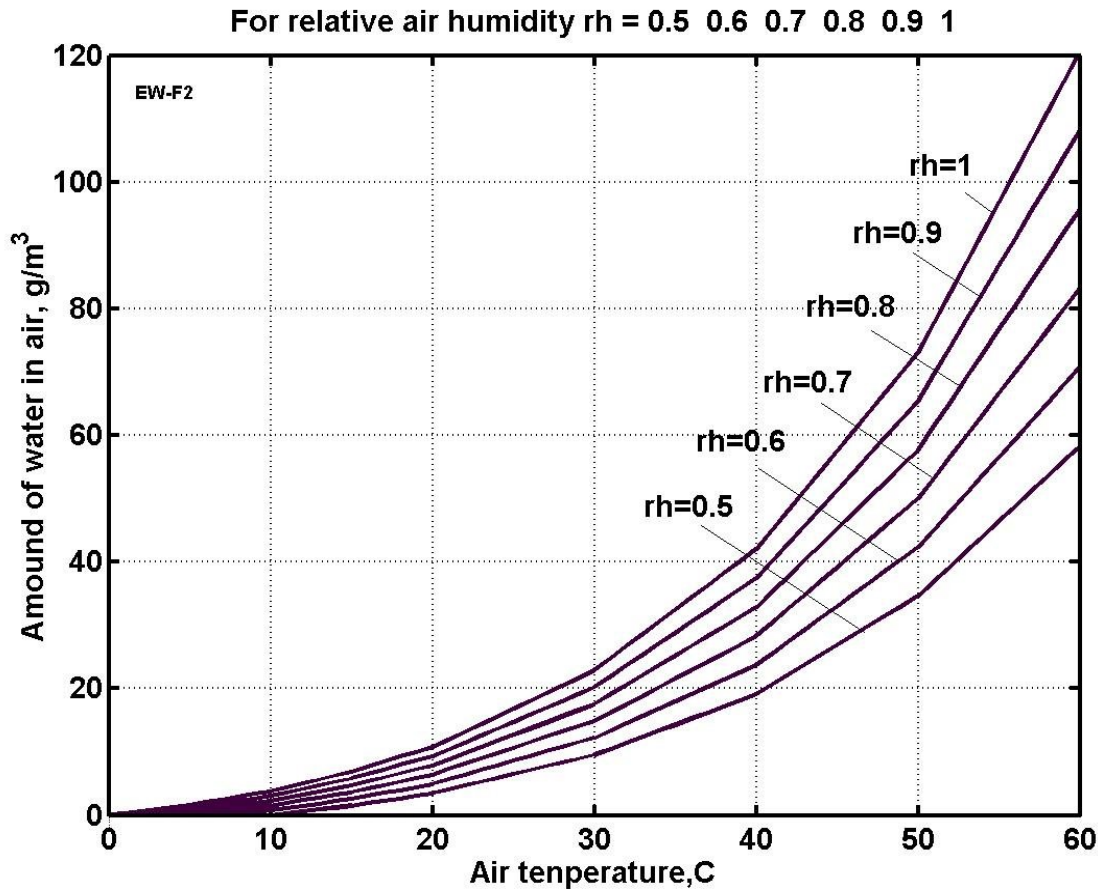

**Fig. 4**. Amount of water in 1 m$^3$ of air versus air temperature and relative humidity (rh). $t_1 = 0$ °C.

Standard atmosphere is in Table 2.

**Table 2**. Standard atmosphere. $\rho_o = 1.225$ kg/m$^3$

| H, km     | 0     | 0.4   | 1     | 2     | 3     | 4     | 5      | 6     |
|-----------|-------|-------|-------|-------|-------|-------|--------|-------|
| t, °K     | 288.2 | 285.6 | 281.9 | 275.1 | 268.6 | 262.1 | 265.6  | 247.8 |
| t, °C     | 15    | 12.4  | 8.5   | 2     | -4.5  | -11   | - 17.5 | -24   |
| ρ/ρ$_o$   | 0     | 0.907 | 0.887 | 0.822 | 0.742 | 0.669 | 0.601  | 0.538 |

2. **The wind dynamic pressure** is computed by equation

$$p_d = \frac{\rho V^2}{2}, \qquad (2)$$

where $p_d$ is wind dynamic pressure, N/m$^2$; $\rho$ is air density, for altitude $H = 0$ the $\rho = 1.225$ kg/m$^3$; $V$ is wind speed, m/s.

The same equation (2) is used for decreasing of air pressure in the tube exit. The computation is presented in fig.5 (Fig.5 is deleted because the size of article for http://arxiv.org is limited 1 Mb) .

3. **Additional decreasing air pressure from warm air**. When entrance air is more warm then atmospheric air (one is heated in solar heater, fig.1) the pressure of tube air column is less the atmosphere pressure and air is sucked by vertical tube. This additional air pressure (rarefaction) may be estimated by equation

$$p_T = p_0 \left[ \exp\left(-\frac{\mu g H}{R \cdot (t + dt)}\right) - \exp\left(-\frac{\mu g H}{R t}\right) \right], \qquad (3)$$

where $p_T$ is additional air pressure (rarefaction), N/m$^2$; $p_0$ is atmospheric pressure on Earth's surface, N/m$^2$; $\mu = 28.96$ is molar weight of air ; $g = 9.81$ m/s$^2$ is Earth gravity; $H$ is altitude, m; $R = 8314$ is gas constant, $t$ is average air temperature, K; $dt$ is increasing of air temperature, K.

The computations are presented in fig.6.

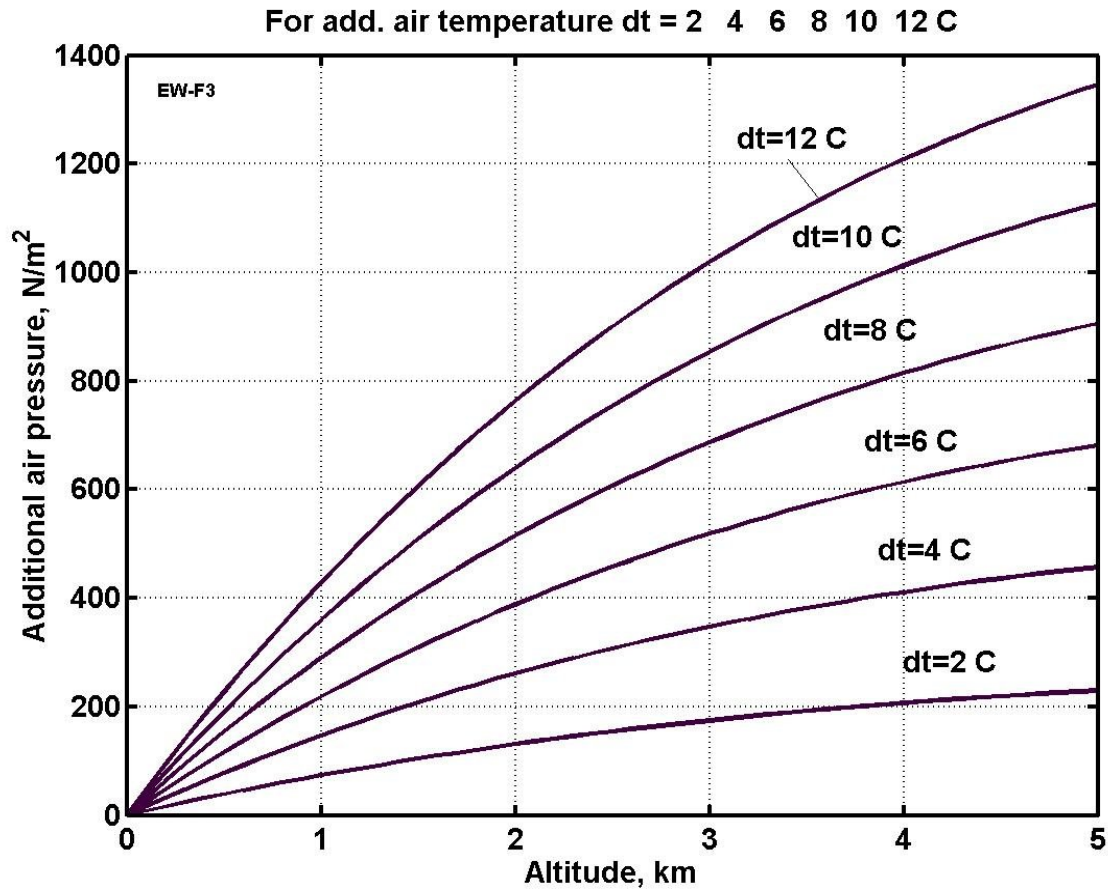

**Fig. 6**. Additional air pressure (rarefaction) into vertical tube versus altitude and additional air warming (dt) by solar heater.

4. **Altitude wind speed**. Wind speed, $V$, increases with altitude, $H$, as follows

$$V/V_o = (H/H_o)^\alpha, \qquad (4)$$

where $\alpha = 0.1 - 0.25$ exponent coefficient depends from surface roughness. When the surface is water, $\alpha = 0.1$; when surface is shrubs and woodlands $\alpha = 0.25$. The sub "0" means the data at Earth surface. The standard values for wind computation are $V_o = 6$ m/s, $H_o = 10$ m/s. The computation of this equation are presented in fig. 7 (Fig.7 is deleted because the size of article for http://arxiv.org is limited 1 Mb).

**5. The air friction of tube walls**. The air friction of tube walls is computed by equation

$$F = C_f \frac{\rho V_t^2}{2} S_f, \qquad (5)$$

where $F$ is friction force, N; $C_f$ is friction coefficient, $C_f = 0.001 \div 0.002$ for laminar air flow and $C_f = 0.005 \div 0.01$ for turbulent air flow; $\rho$ is average air density, kg/m$^3$; $V_t$ is air speed into tube, m/s; $S_f$ is a friction surface, m$^2$.

**6. Air speed into tube**. From balance of pressure we can get the equation for air speed into tube

$$V_t = \sqrt{\frac{2(p_{d,1} + p_{d,2} + p_T)}{\rho(1 + C_f S_f / S)}}, \qquad (6)$$

where $p_{d,1}$, $p_{d,2}$ are wind dynamic pressure in entrance and exit of tube respectively, N/m$^2$; $p_T$ is warm air pressure, N/m$^2$; $\rho$ is average air density into tube, kg/m$^3$; $S$ is cross-section area of tube, m$^2$.

The results of computation are shown in fig. 8.

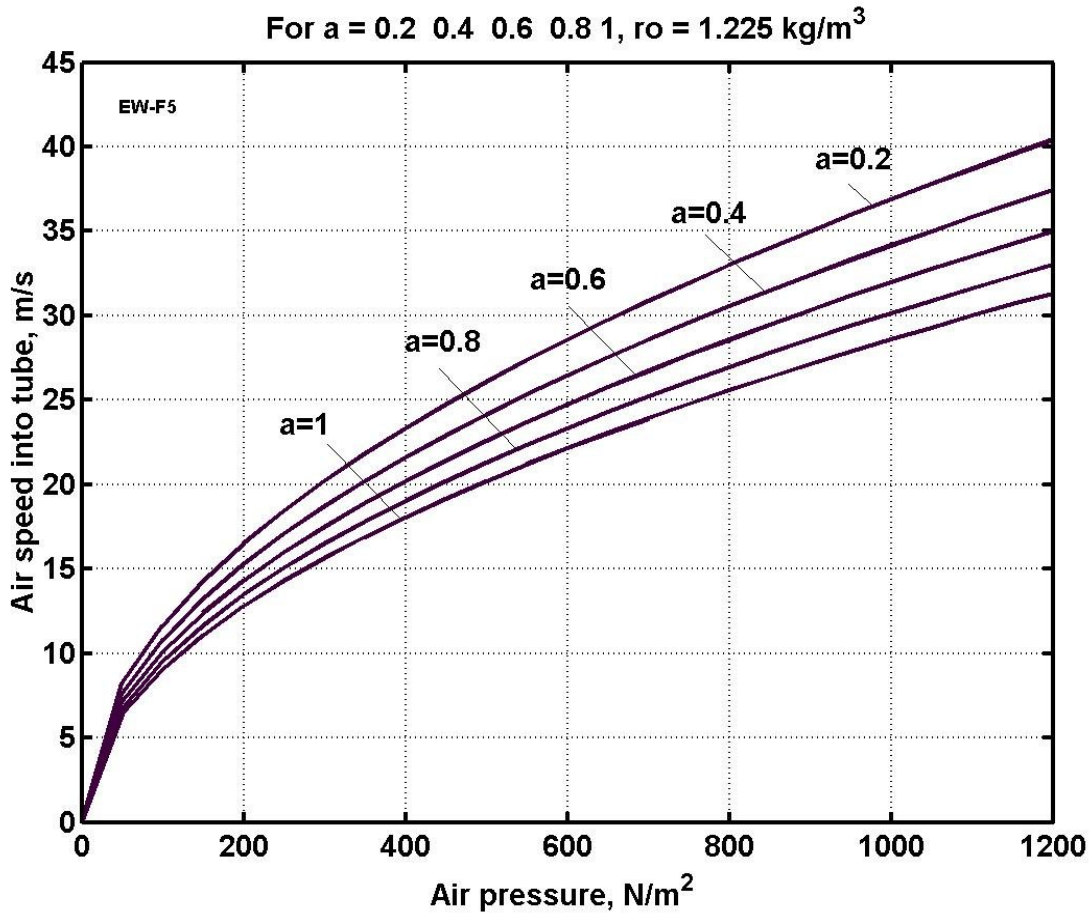

**Fig. 8.** Air speed into tube versus sum of air pressure for different values $a = C_f S_f / S$.

7. **Air propeller at tube entrance** (spending of energy for air pumping). The sea air-water extractor needs the air propeller (pump) in entrance of tube for case when the wind is absent or for the increasing the fresh water production. This propeller may be used when no wind and sun. The power $N$ (and consumption of energy) of this propeller can be computed by equation

$$N = \frac{m_a V_a^2}{2\eta}, \quad m_a = \rho S V_a, \quad N = \frac{\rho S V_a^3}{2\eta}, \tag{7}$$

where $m_a$ is mass of pumped air, kg/s; $V_a$ is additional speed of pumped air, m/s; $\eta \approx 0.8 \div 0.9$ is coefficient of propeller efficiency. The computations are presented in fig.9.

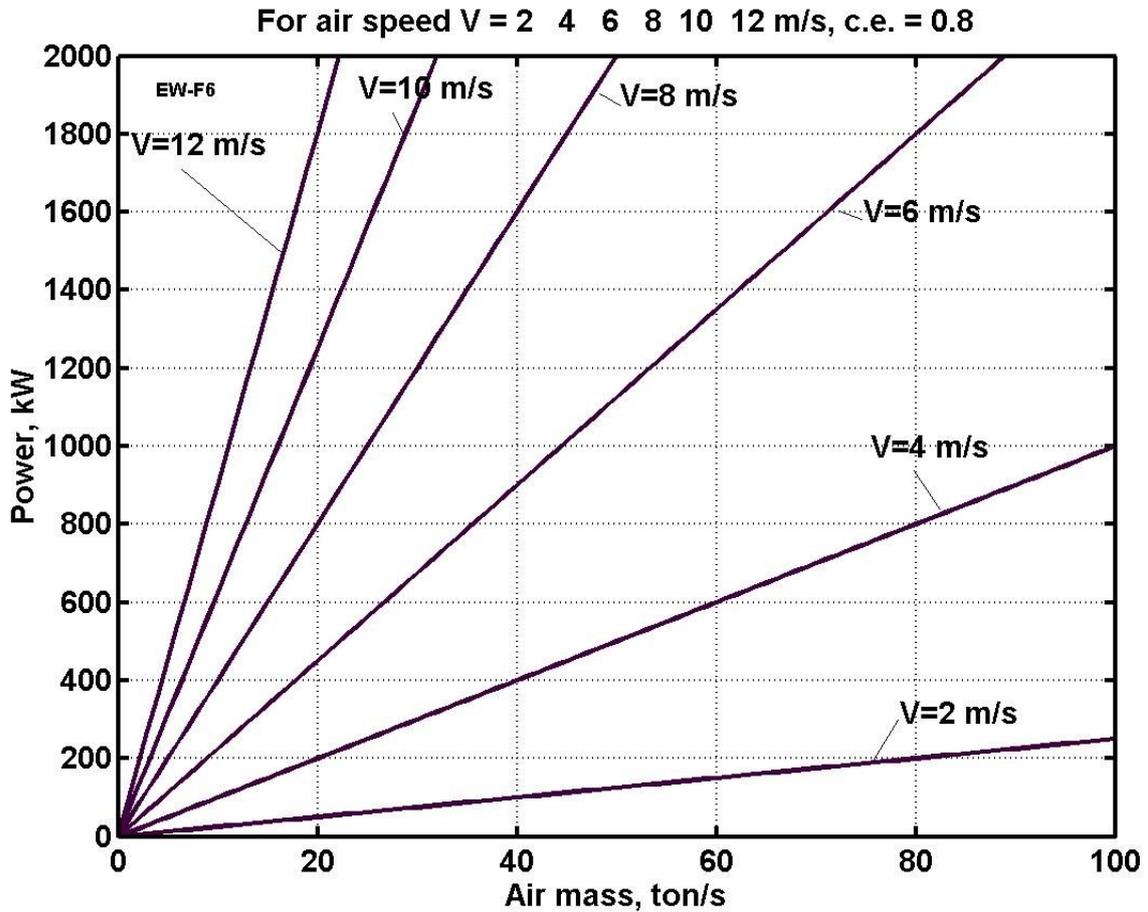

**Fig. 9.** The power needed for air pumping versus air pumped mass and additional air flow speed. Coefficient pump efficiency (c.e.) $\eta = 0.8$.

8. **Getting of energy from high altitude freshwater.** The energy may be received from freshwater condensed into vertical tube at high altitude. The power of high altitude freshwater $N_h$ can be computed by equation

$$N_h = \eta M_w H, \tag{8}$$

where $\eta = 0.9 \div 0.94$ is summary efficiency of water turbine and electric generator; $M_w$ is mass of freshwater, kg; $H$ is altitude of condensation, m.

As an interesting example we compute the energy from high altitude freshwater. Let us take the following initial typical data for air (non sea) extractor: air temperature is t = 35 C, humidity is $h$ = 0.7, the entrance area of vertical tube $S = 4 \times 10^4$ m² (radius tube is 113 m), air speed in tube is $V = 4$ m/s, standard air density $\rho = 1.225$ kg/m³.

Then the air consumption $C = \rho S V = 160,000$ m³/s (about 200 tons/s). One m³ of air contains $m_w = 0.0216$ kg/m³ of freshwater (see Eq. (1)) suitable for condensation. The flow of freshwater (freshwater productivity of installation) is $M_w = m_w C = 3460$ kg/s, Liter/s). They produce an energy $N = \eta M_w H = 31 \times 10^3$ kW.

The reader can say - you get perpetual engine! When wind and Sun are absent (night time), you spend 2000 kW power for moving (blowing) 200 tons/s of air for speed 4 m/s (fig.9) and get 31,000 kW energy (in 15 times more!) plus freshwater. It is not surprise. The air expends at high altitude and produce energy. The same situation is in rain, mountain rivers and glaciers.

**9. Getting of energy from solar heater.** When the vertical tube extractor has a solar air heater at lower end and air turbine (plus electric generator) at top of tube, we can get additional energy. The solar radiation heats the air. The warm air makes a less pressure into tube; the air flow into tube has more (high) speed and air turbine at a top tube end produces the energy. This energy (power) may be estimated by equation

$$N = \eta q S_h, \quad (9)$$

where $N$ is power, W; $\eta$ is efficiency coefficient, $q=1400$ W/m$^2$ is solar power at Earth's orbit in 1 m$^2$, W/m$^2$; $S_h$ is area of solar heater, m$^2$.

The coefficient of efficiency is production of efficiency coefficients of the series devices which take part in energy transferring and change:

$$\eta = \eta_1 \eta_2 \eta_3 \eta_4, \quad (10)$$

where $\eta_1 = 0.1 \div 0.8$ allows for the loss in atmosphere (top value for cloudless time); $\eta_2 = 0 \div 1$ account the heater location and its position to solar ray direction; $\eta_3 = 0.8 \div 0.95$ allows for air friction loss into tube; $\eta_3 = 0.9 \div 0.95$ is common air turbine + electric generator efficiency coefficient. As the result the solar heater has $\eta = 0 \div 0.75$. If solar heater has area 1×1 km, located near equator and sunny day, the maximum energy will be about N = 0.7×1400×10$^6$ ≈ 1 GW energy. The daily energy (include the night time) will be in three-four times less approximately 250 ÷ 330 MW.

If we make transparency the outer layer of tube and black the internal layer, the tube may be used as solar heater. The tube having diameter equals 200 m and height 4 km has useful (for heating) surface area 0.8 km$^2$.

The tube exit air speed $V$ may be estimated by equations

$$N = \frac{M_a V^2}{2}, \quad M_a = \rho V^2 S, \quad V = \sqrt[3]{\frac{2N}{\rho S}}, \quad (11)$$

where $\rho$ is air density at altitude, kg/m$^3$; $S$ is cross-section of tube exit, m$^2$.

**10. Seawater injection into tube.** Correct injection seawater into air flow increases the humidity air (up to maximum), but with other side vaporization of water request a lot of energy and decreases the temperature of air flow. This decreasing of air temperature we can estimate from equation of heat balance

$$rM_w \approx \lambda \rho v \cdot \Delta t, \quad (12)$$

where $r = 2260$ kJ/kg is energy of water vaporization; $M_w$ is expensed mass of water, kg/s; $\lambda = 1$ kJ/kg C is heat capacity of air; $v$ is air flow, m$^3$/s; $\Delta t$ is increasing of air temperature, C.

From equation (12) we find that vaporization 1 g water decreases the temperature of 1 m$^3$ air in 1,84$^o$ C. That decreasing is bad for atmospheric extractor with vertical tube because the decreases the speed of air flow (and total amount of freshwater) and that is good for sea extractor because decreases sea heat exchanger and increases the freshwater production. Last means: the sea exchanger must works all time in maximum humidity of air.

**11. Solar heater for Vertical Extractor.** The size of solar heater for vertical extractor - electric generator may be estimated from heat balance:

$$Q = \lambda M_a \Delta t, \quad M_a = \rho V_t S, \quad S_h = Q/q, \quad (13)$$

where Q is request heat, J; $\lambda = 1$ kJ/kg C is heat capacity of air; $M_a$ is air mass flow, kg/s; $V_t$ is air speed into tube entrance, m/s; $S$ is cross-section area of tube entrance, $S_h$ is requested the heater area, m$^2$; $\rho = 1.225$ kg/m$^3$ if air density; $q \approx 6000 \div 1000$ W/m$^2$ is solar radiation on Earth's surface.

Equation (13) gives: for heating of 1 m$^3$/s of air flow in 1 $^o$C we need a minimum about 1.5 ÷ 2 m$^2$ of the solar heater.

In day time the Sun heats also the heater ground or rather shallow lake (reservoir) with sea water. We ca uses it for heating (humidity) of air in night time.

12. **Heat exchanger for sea extractor**. The transferring of heat in sea exchanger may be computed by equation

$$q_e = k \cdot \Delta t, \quad k = \frac{1}{1/\xi\alpha_1 + \delta/\lambda + 1/\alpha_2}, \qquad (13)$$

where $q_e$ is heat transmission, W/m$^2$; $k$ is coefficient of heat transmission, W/m$^{2\cdot}$C; $\alpha_1 \approx 100$ is coefficient of heat transmission from air to tube wall, W/m$^{2\cdot}$C; $\alpha_2 \approx 5000$ W/m$^{2\cdot}$C is coefficient of heat transmission from water to tube wall, W/m$^{2\cdot}$C; $\delta$ is thickness of tube wall, m; $\lambda \approx 50$ W/m$^{2\cdot}$C is coefficient of heat transmission from through steel wall; $\xi = 1 \div 20$ is coefficient of ribbing the gilled-tube radiator.

For $\xi = 10$, $\delta = 0.01$ m the coefficient $k = 700$ W/m$^{2\cdot}$C.

13. **Cost of freshwater extractor**. The cost of produced freshwater may be estimated by equation

$$C = \frac{C_i/l + M_e + cE_y}{M_{wy}}, \qquad (14)$$

where $C$ is cost of installation; $l$ is live time of installation, years; $M_e$ is annual maintains; $c$ is cost of energy unit; $E_y$ annual expense of energy (receiving of energy has sign minus); $M_{wy}$ is annual amount of received freshwater.

The retail cost of electricity for individual customers is $0.18 per kWh at New York in 2007. Cost of other energy from other fuel is in [8] p.368. Average cost of water from river is $0.49 - 1.09/kL in the USA.

14. **Energy is requested by the different methods.** Below in Table 3 is some data about expense of energy for different methods.

Table 3. Estimation of energy expenses for different methods of freshwater extraction

| No | Method | Condition | Expense kJ/kL | Getting kJ/kL |
|---|---|---|---|---|
| 1 | Evaporation | Expense only for evaporation* | $2.26 \times 10^6$ | 0 |
| 2 | Freezing | Expense only for freezing, c.e. $\eta = 0.3$ | $1 \times 10^6$ | 0 |
| 3 | Sea extractor | Expense only for pumping, $t = 25$ C, $V = 4$ m/s | $1 \times 10^3$ | 0 |
| 4 | High Tube extr. | $t = 35$ C, $h = 0.7$, tube is black | 0 | $30 \times 10^3$ |

* This expense may be decreases in 2 -3 times when the installation is connected with heat or nuclear electric station.

As you see the offered sea extractor decreases the energy expense in 1000 times. The high tube extractor produce freshwater and give a good energy.

15. **Using the high altitude tube of extractor for tourism, communication and wind energy.**
The high altitude tube tower can give good profit from tourism. As it shown in [8] p.93 for 4800 tourists in day and ticket cost $9 the profit will be about $15 million/year.

The profit may be from communication (TV, cell-telephone, military radars, etc). The very good profit will be from high altitude wind electric station [4] (for wind speed $V = 13$ m/s at $H = 4$ km, wind rotor $R = 100$ m the power will be about 20 MW. In reality the wind speed at this altitude is more strong (up 35 m/s), stable, and rotor may to have radius up 150 m. That means the wind power may reach in 30 - 50 times more power.

## Projects
### 1. High tube freshwater and energy Extractor

Let us to make some estimations of the high altitude tube freshwater extractor. Our data is far from optimum. Our aim is to demonstrate the methods of estimation and some possibility offered idea.

Take the radius of inflatable tube $R = 115$ m ($S = 4 \times 10^4$ m$^2$), height of tube $H = 3$ km, air temperature on Earth's surface $t = 25$ °C, air relative humidity $h = 0.7$, wind speed $V = 6$ m/s. From equations and graphs above we get:
- Amount of freshwater into 1 m$^3$ of air is $m_w = 0.0052$ kg/m$^3$ = 5.2 g/m$^3$ [Eq. (1)].
- The average wind speed at altitude $H = 3$ km for $\alpha = 0.15$ is $V = 14$ m/s [Eq. (2)].
- The wind dynamic pressure at $H = 0$ is $p_1 = 22$ N/m$^2$ and the air wind rarefaction at $H = 3$ km is $p_2 = 89$ N/m$^2$ [Eq. (3)].
- For average coefficient of air friction $C_f = 0.005$ and average air density $\rho = 1.1$ kg/m$^3$ the air speed into tube is $V_t = 12.6$ m/s [Eq. (3)].
- The volume and mass of air flow are $v = SV = 5 \times 10^5$ m$^3$/s, $M_a = \rho v = 612$ tons/s.
- The freshwater flow is $M_w = vm_w = 2600$ L/s = 224640 kL/day = 7,862,4000 kL/year (1 kL = 1ton).

**Energy estimation:**
- Power from freshwater ($H = 2500$ m) is 60 MW.
- Power from wind turbine on tube (one at top, $R = 100$ m, $A = \pi R^2$, $\eta = 0.5$, $V = 12.6$ m/s) is $N = 0.5 \eta \rho A V^3 = 14.3$ MW.
- Power from black tube (heating from Sun radiation, $q = 500$ W/m$^2$) is 345 MW.
- Power from solar heater on Earth's surface ($S_h = 1 \times 1$ km, $q = 500$ W/m$^2$) is 500 MW. If solar heater has area $S_h = 2 \times 2$ km the power will be 2000 MW. That is power of powerful electric station.
- If night time and no wind, we can turn on the lower ventilator. For $V_a = 4$ m/s the request ventilator power is 1.57 MW [Eq.(8)]. But the getting energy from high altitude freshwater is 19 MW.

**Cost of high tube extractor-generator.** The cost of thin film (main construction material of inflatable tower and solar heater) is about $0.1/m$^2$ US. The full area of tower ($H = 3$ km) is 2.2 km$^2$, the area of Solar Heater (2×2 km) is 4 km$^2$. The total cost of film is about $0.62 million (mln). Add the cost of 3 ventilator-electric generators - $3 mln. The total cost of offered installation (include building) will be about $10 mln. That is the freshwater extractor of a productivity 224640 kL/day and the powerful electric station with maximal power more then 2000 MW.

The 1 MW of a nuclear electric station costs about $1 mln. The offered installation (electric plant same power) is cheaper a same nuclear electric station in 200 times. One is safety, friendly to environment, and produce free energy and freshwater. The nuclear station requests the nuclear fuel and produce energy which costs as in the conventional heat electric station.

Remainder, the our estimation (data) is not optimal.

### 2. Sea freshwater extractor

Let us take the following initial data for estimation: air temperature is $t = 30°$ C, we have 4 tubes, each has radius R = 5 m, speed of pumped air V = 4 m/s, relative $h = 1$ (after injection of sea water), $m_w = 0.023$ kg/m$^3$.

Then:
- entrance cross-section area is $S = 4 \times \pi \times R^2 = 314$ m$^2$,
- second volume of air flow into tubes is $v = SV = 1256$ m$^3$/s,
- expenses air mass is $M_a = \rho v = 1540$ kg/s,
- produced freshwater is $M_w = vm_w = 35.4$ L/s = 3060 kL/day.
- need (ventilator) power ($\eta = 0.82$) $N = M_a V^2/2\eta = 15$ kW,
- amount of the receding heat (from $t_2 = 30$ C to $t_1 = 10$ C) is $Q = C_p M_a(t_2 - t_1) = 31 \times 10^6$ J/s,
- needed tube area of sea radiator for $k = 700$ W/m$^2$C is $S_r = Q/k = 4.4 \times 10^4$ m$^2$.

If we use the standard steel tubes $d = 1.2$ m, the radiator requests 58 tubes length 200 m.

## Discussing

Author began this research as investigation of new method for receiving the cheap freshwater from atmosphere. In processing research he discovered that method allows to produce huge amount energy, in particular, by transferring the solar energy into electricity with high efficiency (up 80%). If solar cell panels are very expensive and has efficiency about 15%, the thin film (as shopping bags) is very cheap. They are thrown out hundreds of tons every day and waste an environment. The theory of inflatable space towers [1]-[5] allows to build very cheap high height towers, which can be used also for tourism, communication, radio-location, producing wind electricity, space research [8].